\documentclass[prl,twocolumn,groupedaddress,showpacs,nofootinbib]{revtex4}
\usepackage{graphicx}
\usepackage{dcolumn}
\usepackage{bm}
\usepackage{amssymb}
\usepackage{mathrsfs}
\usepackage{amsmath}
\usepackage{epsfig}
\usepackage[dvips]{color}
\usepackage{hhline}
\usepackage{color}

\begin{document}

\title{
Superluminal Neutrinos at OPERA Confront Pion Decay Kinematics }

\author{Ramanath Cowsik$^{1}_{}$}

\author{Shmuel Nussinov$^{2,3}_{}$}

\author{Utpal Sarkar$^{1,4}_{}$}

\affiliation{
$^{1}_{}$Department of Physics and McDonnell Center for the Space
Sciences, Washington University, St.\ Louis, MO 63130, USA\\
$^{2}_{}$ School of Physics and Astronomy, Tel-Aviv University,
Ramat Aviv, Tel Aviv 69978, Israel \\
$^{3}_{}$ Schmidt College of Science, Chapman University,Orange CA 92866, USA\\
$^{4}_{}$  Physical Research Laboratory, Ahmedabad 380009, India}

\begin{abstract}

Violation of Lorentz invariance (VLI) has been suggested as an
explanation of the superluminal velocities of muon neutrinos reported by 
OPERA. In this note we show that the
amount of VLI required to explain this result poses severe difficulties
with the kinematics of the pion decay, extending its lifetime and
reducing the momentum carried away by the neutrinos. We show that
the OPERA experiment limits $\alpha = (v_\nu - c)/c < 4 \times 10^{-6}$.
We then take recourse to cosmic ray data on the spectrum of muons and
neutrinos generated in the earth's atmosphere
to provide a stronger bound on VLI:
$(v-c)/c < 10^{-12}$.

\end{abstract}

\pacs{14.60.Lm 03.30.+p 11.30.Cp 12.60.-i}

\maketitle

The recent OPERA report \cite{opera} of superluminal velocities for the muon
neutrinos, $v(\nu_{\mu})/c = 1+\alpha,~ \alpha=2.5 \times 10^{-5}$, has
evoked much interest. Indeed present information on neutrino
oscillations suggests much stronger bounds on putative
superluminal anomalies for neutrinos \cite{CG,CR}. Still this recent
experiment and  previous measurements at Fermilab \cite{mino1} and
MINOS \cite{minos} supporting this result prompted many
theoretical and phenomenological comments. These possibilities
include speculations of segregating the effect only
into the $\nu_{\mu}$ sector \cite{op1,op2,op3}. In this paper we
study the implications of the superluminal velocities of the
neutrinos on the kinematics of pion decay and show that
superluminal velocities for $\nu_\mu$ are severely constrained by
these considerations. The constraints derived here are not
restricted to any specific model but merely probe into consequences
of superluminal motion of $\nu_\mu$ from pion, kaon and other
decays.

Most of the attempts to understand the OPERA result consider
violation of Lorentz invariance (VLI) at the phenomenological
level \cite{CG,CR,HW,rc,kos,GZK}. 
There are also theoretical motivations
stemming from string theory and from models with extra dimensions.
In these models VLI increases with energy as a power law and have
the general characteristic of modifying the maximum attainable
velocity of the particles. 

The phenomenology of these models have been extensively
studied \cite{CR,rc,kos,GZK}, and important constraints on
the level of VLI have been established. 
Of particular interest is the work of Cohen and Glashow \cite{cohen},
who discuss the possibility of $\nu_\mu \to \nu_\mu + e^- + e^+$
or $\nu_\mu \to \nu_\mu + \nu_e + \bar \nu_e$ and derive strong
constraints on VLI. Other ideas of emission of gravitational
radiation have also been discussed \cite{ellis}. Keeping these in mind,
additional assumptions are required to accommodate the large
superluminal velocities reported by the OPERA collaboration.
The very severe constraints come from the
neutrino sector: neutrino-oscillation experiments suggest that the amount
of Lorentz non-invariant contribution for all the three neutrinos to be the
same ($\alpha_{\nu_e} = \alpha_{\nu_\mu} = \alpha_{\nu_\tau} = \alpha_\nu$),
as noted by Coleman and Glashow \cite{CG,CR}.
The observations of neutrinos from the supernova SN1987A \cite{sn}
require that $|\alpha_\nu| < 10^{-9}$.
The recent OPERA claim of $\alpha_{\nu_\mu} = 2.5 \times 10^{-5}$,
together with the supernova SN1987A constraints seems to indicate that the VLI
parameter grows rapidly with energy, as suggested by some models.

In this paper we note that such a large value of $\alpha_{\nu_\mu}$,
whether energy-independent or energy-dependent, will have many
other phenomenological manifestations. Specifically, they would
affect the kinematics  and the rate of $\pi \to \mu + \nu_{\mu}$
decay, for high energy pions in ways that many experiments (OPERA
included) would have detected. Moreover, the
change in the rate of pion decay would affect the flux of the
cosmic ray muons and muon neutrinos significantly, in conflict
with observations which extend up to $\sim 10^4$ GeV and $\sim
10^5$ GeV, respectively. In the present analysis we assume that
the neutrinos detected at Gran Sasso arise exclusively from pion
decay, even though there could be some contributions from kaons.
As a justification of this assumption we note that the charged kaon
multiplicity at these energies is low $\sim 0.3$, much smaller than
the pion multiplicity of $\sim 6$, and that the transverse momenta
of kaons are larger than that of the pions and the transverse
momenta of neutrinos arising in $K$-decay are larger than those
from $\pi$-decay, so that the probability of $K \to \mu + \nu_\mu$
contributing to the detector at Gran Sasso 730 kms away is 
reduced. Furthermore, the considerations presented here are equally
applicable (with some numerical modifications) to $K$-decay as
well. A more detailed analysis of the kaon contributions is
certainly warranted both in the context of OPERA results and for
the analysis of cosmic-ray data, but will not change the 
conclusions of this paper significantly.

In the formalism for VLI, given by Coleman and Glashow
\cite{CG}, different particles achieve different terminal
velocities, and accordingly, for the discussion of $\pi$-decay,
we make the minimal assumption that
muon neutrinos have superluminal motion and the $\pi^\pm$,
$\mu^\pm$, being charged particles, have terminal velocities
equal to the velocity of light to avoid Cerenkov radiation in vacuum. 
Thus unlike the analysis reviewed in the introduction, our analysis
presented here does not directly apply to $\nu_e$ and $\nu_\tau$,
except indirectly because of neutrino oscillations. However, the
two-body kinematics presented here, with appropriately chosen
$\alpha$, is valid in all cases where one of the emergent particles 
has a superluminal terminal velocity. In models, where $\alpha$ 
increases with energy, the constraints
derived in this paper become much more stringent. 
We make the following standard assumptions:\\[-.225in]
\begin{itemize}
\item[A1] Energy-momentum conservation. \\[-.25in]
\item[A2] The relation $\partial E/{\partial p} =v$ between the velocity of a particle
and the change of its energy with momentum. This classical relation applies also to
the group velocity of waves,
$v_{group} = \partial \omega/{ \partial k}$, and extends it to wave mechanics as well.  \\[-.25in]
\item[A3] The positivity of energy for free particles (by which we exclude Tachyons).
The assumed  $E-p$ relations for different
fields are variants of deformed forward light cones or mass hyperboloids.
\end{itemize} 

\noindent{}\\[-.225in]
These criteria are applicable to most existing VLI models.

The assumption A2 for the muon neutrinos implies
\begin{equation}
 \int dE = \int v(p) dp \,,
\end{equation}
where $v(p)>1$ beyond some small value of momentum $p_{min}$ that is 
much larger than the tiny sub eV mass of the neutrinos, $m_\nu$, which
we neglect. 
Defining, in general,
\begin{equation}
\left\langle \frac{\partial E}{ \partial p} \right\rangle = 1 + \alpha
\,,
\end{equation}
as the effective average over the neutrino momenta detected in 
the OPERA experiment, centered around 17 GeV, 
we write the energy-momentum relation at high energies as
\begin{equation}\label{3}
 E_\nu = p_\nu ( 1 + \alpha) \,,
\end{equation}
where $\alpha$ corresponds to VLI, required to understand the
OPERA anomaly.
The kinematic analysis begins with the standard mass-energy relation
for $\pi,~\mu$:
\begin{equation}\label{4}
 E_i= (p_i^2+m_i^2)^{1/2} \,.
\end{equation}
We then express the four vector of the decaying pion as
\begin{equation}\label{5}
 \left( E_\pi, p_\pi, 0 , 0 \right)
\end{equation}
and those of the final neutrino and muon as
\begin{equation}\label{6}
 \left( E_\nu, p_{\nu \ell}, p_{\nu t} , 0 \right)  ~~ {\rm and} ~~
 \left( E_\mu, p_{\mu \ell}, p_{\mu t}, 0 \right),
\end{equation}
respectively, where the longitudinal components of momenta are taken to be
$p_{\nu \ell} = \eta p_\pi$ and $p_{\mu \ell} = (1-\eta) p_\pi$ and the
transverse components as $p_{\nu t} = -p_{\mu t} = p_t$.
With this choice the conservation
of all the spatial components of momenta is evident.

We still need to satisfy the energy conservation:
\begin{equation}\label{7}
 E_\pi = E_\nu + E_\mu \,,
\end{equation}
with: {
\begin{eqnarray}\label{8}
 & E_\pi = \left[ p_\pi^2 + m_\pi^2 \right]^{1/2}; 
 ~~ E_\nu = \left[ p_\pi^2  \eta^2 + p_t^2 \right]^{1/2}(1 + \alpha)& \nonumber \\
  &{\rm and}  ~~~~~~ E_\mu = \left[ p_\pi^2 (1 - \eta)^2 + p_t^2 + m_\mu^2 
  \right]^{1/2} \,.
\end{eqnarray}
}

Keeping in mind that in accelerator experiments including OPERA,
$m_\pi/p_\pi$, $m_\mu/p_\pi$ and $p_t/p_\pi$ are
very small, we expand the square root and keep only the leading term
to get
\begin{eqnarray}\label{9}
 \frac{m_\pi^2}{2 p_\pi} &=& \frac{p_t^2 + m_\mu^2}{2 p_\pi(1 - \eta)}
 + \eta \alpha p_\pi + \frac{p_t^2(1 + \alpha)}{2 p_\pi \eta} \,.
\end{eqnarray}
Rearranging we can write
\begin{equation}\label{10}
 \frac{m_\pi^2}{2 p_\pi^2} - \frac{m_\mu^2 + p_t^2}{2 p_\pi^2 (1 - \eta)}
 - \frac{p_t^2}{2 p_\pi^2 \eta} = \alpha \left( \eta + \frac{p_t^2}{2 p_\pi^2 \eta}
 \right)
\end{equation}
{ \vskip -.3in
\begin{equation}\label{11}
{\rm or}~~~ \alpha = \frac{1}{2 p_\pi^2 \eta^2 + p_t^2} \left[m_\pi^2 \eta - m_\mu^2
 \frac{\eta}{(1 - \eta)}
 - p_t^2 \frac{1}{(1-\eta)}  \right] \,.
\end{equation} }
Since $p_t^2$ is positive, this yields a constraint:
\begin{equation} \label{12}
 \alpha \eta \leq \frac{1}{2 p_\pi^2 } \left[ m_\pi^2 - \frac{m_\mu^2}{(1-\eta)} \right] \,.
\end{equation}
In the OPERA experiment the typical energy of the neutrinos is
$\sim 17$ GeV that arise from the decay of pions with a mean
energy of $\sim 60$ GeV, so that the typical value of $\langle
\eta \rangle \approx 0.3$. Inserting this value of $\eta$ into
equation \ref{12} we obtain the bound:
\begin{equation}\label{13}
  \alpha_{OPERA} \leq \frac{1}{0.6 p_\pi^2} \left[ m_\pi^2 -
  \frac{m_\mu^2}{0.7} \right] \sim 4 \times 10^{-6} \,.
\end{equation}
Note that this bound on the superluminal parameter, $\alpha$, is
significantly smaller than $2.5 \times 10^{-5} $ estimated from
the time profiles of the events and the GPS timing in their
experiment.

Next, we address the question, whether $\eta$ could indeed be
smaller than the assumed value of $\sim 0.3$ which would allow the
value of $\alpha$, estimated in our analysis, to be consistent
with the OPERA result. For this, special selection effects should
conspire to push $\eta$ down to $\sim 0.05$. We note that this
hypothesis would imply significant enhancement of the lifetime of
the pions. To see this, note that within 
this standard kinematics of pion decay, the
value of the $\eta$ parameter is uniformly distributed in the
range 
\begin{equation}
 0 \leq \eta \leq 1 - \frac{m_\mu^2 }{m_\pi^2} \,,
\end{equation} 
{\it i.e.}, in the range $\sim 0 - 0.5$. The phase space for the
pion decay is directly proportional to this range and any
reduction in this range will have a corresponding reduction in the
rate of decay. It is straightforward to perform the Lorentz 
non-invariant phase space integral after modifying the 
$\delta$-functions representing mass shell conditions
according to equations \ref{3}--\ref{8}. Such a calculation
yields an integral directly proportional to $\eta_{max}$.
Thus with the reduction of $\langle \eta \rangle$ 
to $0.05$, the pion lifetime will be extended by a factor
of 6 or more, which is excluded by various accelerator experiments,
including OPERA.

As seen clearly from equations \ref{10}-\ref{12}, the bounds get stronger
in proportion to $p_\pi^2$, or even with higher powers in models
with $\alpha$ increasing with energy invoked recently for explaining
how the the OPERA results need not be flavor
specific and still be consistent with the small $\alpha$ inferred
from SN1987A neutrinos. Accordingly the bounds on VLI become 
extremely stringent for the
ultra-high energy muons and neutrinos observed in deep underground
experiments at Kolar Gold Fields, Kamiokande, Baksan, Ice-Cube and
other experiments \cite{ice,baksan,ice2}.

Before we discuss these cosmic ray observations, we note that the
fraction of the momentum carried away by the muon in the standard
decay kinematics of the pion, $(1-\eta)$ is in the range of $\sim 0.5-1$.
The spectrum of muons generated in the Earth's atmosphere is well
measured up to energies of $\sim 10^5$ GeV and we confine our
analysis to the spectrum up to $\sim 4 \times 10^4$ GeV where the
muons arise mainly from the decay of pions and kaons and the
contributions of muons generated by neutrino interactions
in rock to the depth intensity curve could be neglected. The observed
differential energy spectrum is well represented by the
theoretical estimate \cite{cptv}:
\begin{eqnarray}\label{16}
 f_\mu(E)& \cong & \left[ A_\pi \langle 1 - \eta \rangle_\pi^{\beta-1}
 \left\{ \frac{ \langle 1 - \eta \rangle_\pi {\cal E}_\pi}{E + \langle 1 - \eta
 \rangle_\pi {\cal E}_\pi} \right\} \right. \nonumber \\
 &+& \left . A_K \langle 1 - \eta \rangle_K^{\beta-1}
 \left\{ \frac{ \langle 1 - \eta \rangle_K {\cal E}_K}{E + \langle 1 - \eta \rangle_K 
 {\cal E}_K } 
 \right\} \right] E^{- \beta} \,,\phantom{xxx}
\end{eqnarray}
where: \\ {
\noindent 
$\beta =$ spectral index of the cosmic ray spectrum $\sim
 2.65$ \\
${\cal E}_{\pi} = h_0 (\theta) m_{\pi } / c
 \tau_{\pi}$ \\
${\cal E}_{K} = h_0 (\theta) m_{K } / c
 \tau_{K}$ \\
$h_0(\theta)= 7 \times 10^{5} \sec \theta$ cm, the scale
 height of the Earth's atmosphere at a zenith angle $\theta$ \\
$\tau_{\pi/K}=$ decay lifetimes of pions/kaons at
 rest \\
$\langle 1 - \eta \rangle_{\pi /K} =$ the fractional
 momenta carried away by the muons in pion/kaon decay averaged
 over the spectrum of cosmic rays, around the energy band of
 interest \\
$A_{\pi /K}=$ Constants.
}

These constants are estimated from the inclusive cross sections
for the production of pions and kaons at high energies and indicate that the net
contribution of $K$-decay is $\sim 10\%$ for the muons and about
$\sim 70\%$ of the total flux of neutrinos at the highest energies. A
similar expression for the flux of neutrinos generated in the
atmosphere results when we replace $\langle 1 - \eta \rangle$ by
$\eta$ in equation \ref{16}. Notice that at very high energies $\gtrsim 10^3$ GeV,
with $E\gg {\cal E}_{\pi /K}$, the spectra of muons and neutrinos
become steeper with a spectral index $\sim (\beta + 1)$.
Furthermore, the spectral intensities became proportional to $\langle
1 - \eta \rangle^\beta$ or $\langle \eta \rangle^\beta$ as the case might be.

Now the spectrum of muons presented by Novoseltsev et al \cite{baksan}, fits
well with equation \ref{16}, that assumes that $\tau_{\pi /K}$ are
constants. Thus $\langle \eta \rangle$ has to be constant up to energies
of $\sim 4 \times 10^4$ GeV. Note that equation \ref{16} is
sensitive to change in $\langle \eta \rangle$ in two ways -- first
through the change $\langle 1 - \eta \rangle^\beta$ and more
importantly through its effect on extending the lifetime of pions
and kaons. Thus the muon data imply
\begin{equation}\label{17}
    \alpha \eta < 10^{-11} \,.
\end{equation}
{Much more extensive data of the atmospheric muons ($2 \times 10^{10}$ events)
and upward neutrinos ($2 \times 10^4$ events) of energies in the range of $1-400$ TeV,
generated by energetic cosmic rays from the other side of the 
Earth have been provided recently by the south pole Ice-Cube experiment
\cite{ice,ice2}, 
which} shows a good fit
with an index $\sim (\beta + 1) \sim 3.65$ at energies $E \gg
{\cal E}_{\pi /K}$. Thus their observations imply a constraint
\begin{equation}\label{18}
    \alpha \eta < 10^{-13} \,.
\end{equation}
Keeping in mind that we can not allow significant changes in
$\eta$ as they will affect the spectral slope and spectral
intensities of the muons and neutrinos, the limits derived here
represent bounds on the superluminal parameter $\alpha$, which may
be stated conservatively as
\begin{equation}\label{17}
    \alpha  < 10^{-12} \,,
\end{equation}
allowing nearly a factor of ten variance 
for any contributions of the uncertainties in the cosmic-ray
observations and the approximations used in our analysis. 
It should be noted that since spectra of both the muons and the neutrinos 
are fit very well the estimates, which assume $\tau_\pi$ and $\eta$
to be constants, the bound on the  VLI parameter $\alpha$ follows 
exclusively from kinematic considerations.
Indeed accurate spectra of atmospheric muons and/or neutrinos
even at lower energies of a several TeV can be used to improve 
the limits presented here. Our limit on $\alpha_{\nu_\mu}$ is far more 
stringent than that derived from the observations of
the neutrinos emitted in SN1987A. It may be appropriate to note here that
the observation of even a single event initiated by the GZK
neutrinos in detectors like ANITA \cite{anita} would improve
the bound on $\alpha$ to $\sim 10^{-20}$.
To assess the impact of these results on specific models, we note that 
in general the matrix element in VLI theories may also have
novel energy dependence, however they are unlikely to exactly cancel
the above purely kinematic effects derived in this paper.

We would like to draw attention to an independent analysis of the Ice-Cube data
by Bi et al \cite{bi}, who assume that the superluminal $\nu_\mu$
may be treated as having an effective mass, $m_{eff} = [m_\nu^2 + 2 
\alpha p_\nu^2]^{1/2}$, so that the decay mode $\pi \to \mu +
\nu_\mu$ becomes forbidden beyond a threshold energy for the neutrinos. This 
analysis yields result similar to our results, which we have
derived showing the progressive kinematic restriction of the
phase space available for $\pi$-decay, leading to a monotonic
increase of pion life-time with energy.

In summary, we presented here a strong constraint $(v-c)/c < 10^{-12}$ on the
amount of violation of Lorentz invariance from pion decay
kinematics and cosmic ray data. 
Careful observations of the fluxes of very high energy muons and neutrinos at
accelerators and in cosmic rays, and their comparison with the expected
fluxes will constrain any possible variation of the decay life-time
of the pion, which in turn, will lead to better bounds than those
reported here. 

\vspace{5mm}

\noindent \textbf{Acknowledgement}: { One of us (SN) thanks E. Blaufus, G. Sullivan 
and J. Goodman for discussions of the ice-Cube data.} 
US thanks the Washington
University in St.\ Louis for the invitation as the Clark Way Harrison
visiting professor.


\begin{thebibliography}{99}

\bibitem{opera} {\it OPERA Collaboration}: T. Adam, et al, arXiv:1109.4897v1 [hep-ex].

\bibitem{CG} S. Coleman and S.L. Glashow, Phys. Rev. {\bf D 59},
116008 (1999).

\bibitem{CR} S. Coleman and S.L. Glashow, Phys. Lett. {\bf B 405}, 249 (1997).

\bibitem{mino1} G.R. Kalbfleisch, N. Baggett, E.C. Fowler, J. Alspector, Phys. Rev. Lett. {\bf 43}, 1361 (1979);
J. Alspector, et al, Phys. Rev. Lett. {\bf 36}, 837 (1976).

\bibitem{minos}
{\it MINOS Collaboration}: P. Adamson, et al, Phys. Rev. {\bf D 76}, 072005 (2007).

\bibitem{op1} R.B. Mann and U. Sarkar, arXiv:1109.5749 [hep-ph] (2011);
S. Hannestad and M.S. Sloth, arXiv:1109.6282 [hep-ph] (2011);
I.Ya. Aref'eva, and I.V. Volovich, arXiv:1110.0456 [hep-ph] (2011).

\bibitem{op2} G.F. Giudice, S. Sibiryakov, and A. Strumia, arXiv:1109.5682 [hep-ph]
(2011);  G. Amelino-Camelia, G. Gubitosi, N. Loret, F. Mercati, G. Rosati,
and P. Lipari, arXiv:1109.5172 [hep-ph] (2011);
G. Cacciapaglia, A. Deandrea, and L. Panizzi, arXiv:1109.4980 [hep-ph] (2011);
N.D. Haridass, arXiv:1110.0351 [hep-ph] (2011); T. Li and D.V. Nanopoulos,
arXiv:1110.0451 [hep-ph] (2011); H. Gilles, arXiv:1110.0239 [hep-ph] (2011).

\bibitem{op3} A. Drago, I. Masina, G. Pagliara, and R. Tripiccione,
arXiv:1109.5917 [hep-ph] (2011);  F. Tamburini, and M. Laveder,
arXiv:1109.5445 [hep-ph] (2011);  F.R. Klinkhamer, arXiv:1109.5671 [hep-ph]
(2011); L. Gonzales-Mestres, arXiv:1109.6630 [physics.gen-ph] (2011);
P. Wang, H. Wu, and H. Yang, arXiv:1110.0449 [hep-ph] (2011);
arXiv:1109.6930 [hep-ph] (2011).

\bibitem{HW} M.P. Hagen and C.M. Will, Phys. Today {\bf 40}, 69 (1987);
Y. Grossman, C. Kilic, J. Thaler, and D.G.E. Walker, Phys. Rev. {\bf D 72}, 125001 (2005); J. Alfaro, Phys. Rev. Lett. {\bf 94}, 221302 (2005).

\bibitem{rc} R Cowsik and B.V. Sreekantan, Phys. Lett. {\bf B 449}, 219 (1999).

\bibitem{kos} V.A. Kostelecky and N. Russel, Rev. Mod. Phys. {\bf 83}, 11 (2011);
  T.~Hambye, R.~B.~Mann, U.~Sarkar,
  Phys.\ Rev.\  {\bf D58}, 025003 (1998).
  [hep-ph/9804307];
  T.~Hambye, R.~B.~Mann, U.~Sarkar,
  Phys.\ Lett.\  {\bf B421}, 105 (1998).
  [hep-ph/9709350];
V.A. Kostelecky, Phys. Rev. Lett. {\bf 80}, 1818 (1998);
Phys. Rev. {\bf D 61}, 016002 (1999);
R. Bluhm, V.A. Kostelecky, and C.D. Lane, Phys. Rev. Lett. {\bf 84}, 1098 (2000);
H. Pas, S. Pakvasa, T.J. Weiler, Phys. Rev. {\bf D 72}, 095017 (2005);
S. Hollenberg, O. Micu, H. Pas, and T.J. Weiler, Phys. Rev. {\bf D 80}, 093005 (2009).

\bibitem{GZK} S. Coleman and S.L. Glashow, hep-ph/9808446 (1998);
F.W. Stecker and
S.T. Scully, Astroparticle Physics {\bf 23}, 203 (2005);
L. Maccione, A.M. Taylor, D.M. Mattingly, and S. Liberati, JCAP {\bf 0904}, 022 (2009).

\bibitem{cohen} A.G. Cohen, and S.L. Glashow, Phys. Rev. Lett. {\bf 107}, 181803 (2011).

\bibitem{ellis} J. Alexandre, J. Ellis and N.E. Mavromatos, arXiv:1109.6296 [hep-ph] (2011).

\bibitem{sn} L. Stodolsky, Phys. Lett. {\bf B 201}, 353 (1988);
K. Hirata, et al, Phys. Rev. Lett. {\bf 58}< 1490 (1987);
R.M. Bionta, et al, Phys. Rev. Lett. {\bf 58}, 1494 (1987);
M.J. Longo, Phys. Rev. {\bf D 36}, 3276 (1987);
D. Fargion, arXiv:1109.5368 [astro-ph.HE] (2011).

\bibitem{ice} {\it Ice-Cube Collaboration}: J.K. Becker et al,
Nucl. Phys. Proc. Suppl. {\bf 217}, 269 (2011);
{\it BAKSAN and EAS-TOP Collaboration}: M. Aglietta et al, Astropart. Phys.
{\bf 14}, 189 (2000);
F.E. Gray, C. Ruybal, J. Totushek, D.-M. Mei, K. Thomas, and
C. Zhang, Nucl. Instrum. Meth. {\bf A 638}, 63 (2011);
{\it SNO Collaboration}: B. Aharmim, et al, Phys. Rev. {\bf D 80}, 012001 (2009).

\bibitem{baksan} Yu.F. Novoseltsev, A.G. Bogdanov, R.P. Kokoulin, R.V. Novoseltseva,
V.B. Petkov, and A.A. Petrukhin, 
{\it 22nd European Cosmic Ray Symposium}, Turku, Finland, August 3--6, (2010). 

\bibitem{ice2} {\it Ice-Cube Collaboration}: R. Abbasi et-al, Phys. Rev. {\bf D 83}, 012001 (2011).

\bibitem{cptv} R. Cowsik, Yash Pal, S.N. Tandon, and R.P. Verma,
{\it Proc. Ind. Acad. Sciences} {\bf 63}, 217 (1966).

\bibitem{anita} S. Hoover et al, Phys. Rev. Lett. {\bf 105}, 151101 (2010).


\bibitem{bi} X.-J. Bi, P.-F. Yin, Z.-H. Yu, and Q. Yuan, arXiv:1109.6667 [hep-ph] (2011).

\end{thebibliography}
\end{document}